\documentclass[12pt]{article}
\large

\title{Quantum restoration of broken symmetries}

\vspace{0.5cm}

\author{ V.V. Belokurov$^{1,2}$ and E.T. Shavgulidze$^{1}$    \\
\\    {\em 1. Lomonosov Moscow State University, Russia }
\\    {\em 2. Institute for Nuclear Research of Russian Academy of Sciences, Russia }
\\ {\it  belokur@rector.msu.ru ; shavgulidze@bk.ru}}

\date{ \ \ \  }
\begin{document}
\maketitle

\begin{abstract}

A certain non-linear non-local substitution is shown to transform the action of the self-interacting quantum field to the free one. The functional integrals in both theories are equal to each other. However, the integrations are performed over different functional spaces. The classical action and the classical limit of the corresponding quantum theory turn out to be different. And the symmetry originally broken in the classical theory is restored in the classical limit of the quantum theory.

\end{abstract}

\vspace{2cm}

In our previous paper \cite{(BSh)} we have considered the model of $\phi^{4}$ self-interacting quantum field in one-dimensional "space-time". Under the substitution
\begin{equation}
   \label{1}
\chi(t)=\phi(t)+\int \limits _{0}^{t}\phi^{2}(\tau)d\tau
\end{equation}
it turns into the free field model. And we get the formal equality of the functional integrals
\begin{equation}
   \label{2}
\int\,\exp\left\{ -\frac{1}{2}\int \limits _{0}^{1}(\dot{\phi}(t))^{2}dt -\frac{1}{2}\int \limits _{0}^{1}\phi^{4}(t)dt- \int \limits _{t=0}^{t=1}\phi^{2}(t)d\phi \right\}\ d\phi =
\end{equation}
$$
\int\,\exp\left\{ -\frac{1}{2}\int \limits _{0}^{1}(\dot{\chi}(t))^{2}dt \right\}\ d\chi \,.
$$

In eq. (\ref{2}), the last term in the exponent is the Ito stochastic integral (see e.g. \cite{(Ito)})
\begin{equation}
   \label{3}
 \int \limits _{t=0}^{t=1}\phi^{2}(t)d\phi =\frac{1}{3}\left[\phi^{3}(1)- \phi^{3}(0)\right]- \int \limits _{0}^{1}\phi(t)dt\,.
 \end{equation}
Note that Ito integrals appear even at the simplest transformations of the Wiener measure \cite{(Kuo)}.

In \cite{(BSh)}, we have demonstrated that the functions $\chi(t)$ and $\phi(t)$ belong to different functional spaces. Namely, if  $\chi(t)\in C([0,1])\,,$ than $\phi(t)\in X\neq C([0,1])\,.$ Here, $X$ is the space of functions that can have a finite number of singularities of the type $(t-t_{i}^{\ast})^{-1}$ on the interval $[0,1]\,.$  It is quite natural as for a bounded function $\chi(t)$ there can be some regions where $|\varphi(t)|$ becomes large enough. And in these regions the behavior of the function $\phi(t)$ is prescribed by the equation $\dot{\phi}=-\phi^{2}\,.$

Thus, the equivalence of the two theories manifests itself in the equality of the functional integrals
$$
Z=\int\limits _{C[0,1]}\,\exp\left\{ -\frac{1}{2}\int \limits _{0}^{1}(\dot{\chi}(t))^{2}dt \right\}\ d\chi =
$$
\begin{equation}
   \label{4}
\int\limits _{X}\,\exp\left\{ -\frac{1}{2}\int \limits _{0}^{1}(\dot{\phi}(t))^{2}dt -\frac{1}{2}\int \limits _{0}^{1}\phi^{4}(t)dt-
\frac{1}{3}\left[\phi^{3}(1)- \phi^{3}(0)\right]+ \int \limits _{0}^{1}\phi(t)dt \right\}\ d\phi \,.
\end{equation}

We see that for interacting quantum fields the functional space appears to be more singular than the one for the free field. On the space $X\,,$
$
\exp\left\{ -\frac{1}{2}\int \limits _{0}^{1}(\dot{\phi}(t))^{2}dt \right\}\,d\phi
$
is not a measure. An attempt to expand the $\exp$ in powers of the interaction $\phi^{4}$ leads to ambiguous infinite expressions.
 And only
$$
  Z^{-1}\ \exp\left\{ -\frac{1}{2}\int \limits _{0}^{1}(\dot{\phi}(t))^{2}dt -\frac{1}{2}\int \limits _{0}^{1}\phi^{4}(t)dt- \frac{1}{3}\left[\phi^{3}(1)-\phi^{3}(0)\right]+\int \limits _{0}^{1}\phi(t)dt\right\}\ d\phi
$$
can be considered as a measure on $X\,.$

In what follows we will not write the normalizing factors at functional integrals. That is, $Z$ will be set equal to 1.

The similar picture takes place for four-dimensional space-time where the interacting quantum field is not a continuous function but a distribution. The Gaussian measure of the set of continuous functions is equal to zero
$$
\int\limits _{C}\,\exp\left\{ -\frac{1}{2}\int \limits _{\mathbf{R}^{D}}\,\phi(x)\Delta_{(D)}\phi(x)d^{D}x \right\}\ d\phi =0\,,\ \ D=2,\ldots\,,
$$
and the well-known Haag theorem (see e.g. \cite{(Str)}) is valid.

The non-linear non-local substitution (\ref{1}) and the equality of the functional integrals (\ref{4}) lead to an interesting effect. The classical action and the classical limit of the corresponding quantum theory turn out to be different.

First, let us describe the model. For more clearness we slightly modify the interaction term and consider the action
\begin{equation}
   \label{5}
 A=\frac{1}{2}\int \limits _{-T}^{+T}(\dot{\varphi}(t))^{2}dt +\frac{a^{2}}{2}\int \limits _{-T}^{+T}\,\left(\varphi^{2}(t)-\frac{1}{4}\frac{b^{2}}{a^{2}}\right)^{2}dt \,.
\end{equation}

The potential
\begin{equation}
   \label{6}
 V(\varphi(t))=\frac{1}{2}a^{2}\left(\varphi^{2}(t)-\beta^{2}\right)^{2}\ \ ( \beta\equiv \frac{b}{2a}\,;\ a>0,\  b>0 )
\end{equation}
has two degenerate minimae at $\varphi=\pm\beta$ and  a local (unstable) maximum at $\varphi=0\,,$ and is symmetric: $ V(-\varphi )=V(\varphi )\,. $

The Euler-Lagrange equation has the form
\begin{equation}
   \label{7}
\ddot{\varphi}(t))-2a^{2}\varphi\left(\varphi^{2}(t)-\beta^{2}\right)=0\,.
\end{equation}
So, the classical system given by the action $A$ is symmetrical under the substitution $\varphi\rightarrow -\varphi\,.$

For the classical system given by the action
\begin{equation}
   \label{8}
 A_{+}= A -a \int \limits _{-T}^{+T}\varphi(t)dt+\frac{a}{3}\left[\varphi^{3}(+T)-\varphi^{3}(-T) \right]-a\beta^{2}\left[\varphi(+T)-\varphi(-T) \right]
\end{equation}
the symmetry is broken because of the term linear in $\varphi$ and the boundary terms.

The action $A_{+}$ leads to the Euler-Lagrange equation
\begin{equation}
   \label{9}
\ddot{\varphi}(t))-2a^{2}\varphi\left(\varphi^{2}(t)-\beta^{2}\right)+a=0
\end{equation}
and the boundary conditions
\begin{equation}
   \label{10}
\dot{\varphi}(\pm T)+a\left(\varphi^{2}(\pm T)-\beta^{2}\right)=0\,.
\end{equation}

The corresponding quantum theory deals with the functional measure
$$
\int\exp\{-A_{+}(\varphi)\}d\varphi\,.
$$

As it follows from \cite{(BSh)}, the substitution
\begin{equation}
   \label{11}
\chi(t)=\varphi(t)+a\int \limits _{-T}^{t}\left(\varphi^{2}(\tau)-\beta^{2}\right)d\tau\
\end{equation}
results in the equality of the functional integrals
\begin{equation}
   \label{12}
\int\limits _{X^{+}}\,F(\varphi)\,\exp\{-A_{+}(\varphi)\}d\varphi=\int\limits _{C[-T,+T]}\,F(\varphi(\chi))\,\exp\left\{ -\frac{1}{2}\int \limits _{-T}^{+T}(\dot{\chi}(t))^{2}dt \right\}\ d\chi \,.
\end{equation}
The functional space $X^{+}$ can be determined in a similar way as it is done in \cite{(BSh)}. It is the space of functions that can have singularities on the interval $[-T,+T]\,.$

Note that  the term linear in $\varphi$ and breaking the symmetry appears in the integrand of the left-hand side of eq. (\ref{12}) from the Ito integral (see eq. (\ref{3})).
The similar effect occurs in the quantum theory of a particle in a magnetic field  \cite{(Sim)}.

Now consider the classical limit of eq. (\ref{12})
$$
F(\tilde{\varphi})=\lim\limits_{\hbar\rightarrow 0}\int\limits _{X^{+}}\,F(\varphi)\,\exp\{-\frac{1}{\hbar}\,A_{+}(\varphi)\}d\varphi=
$$
\begin{equation}
   \label{13}
   \lim\limits_{\hbar\rightarrow 0}\int\limits _{C[-T,+T]}\,F(\varphi(\chi))\,\exp\left\{ -\frac{1}{\hbar}\,\frac{1}{2}\int \limits _{-T}^{+T}(\dot{\chi}(t))^{2}dt \right\}\ d\chi \,.
\end{equation}

In the right hand side of eq. (\ref{13}) the classical limit yields the equation $\ddot{\chi}(t)=0\,,$ or $\dot{\chi}(t)=const\,.$ Actually, $const=0$ as $\chi(+T)$ is not fixed. It follows from the boundary conditions (\ref{10}) as well.
In terms of the function $\varphi(t)$ the equation looks like
\begin{equation}
   \label{14}
\dot{\varphi}(t)+a\left(\varphi^{2}(t)-\beta^{2}\right)=0\,.
\end{equation}

  Eq. (\ref{7}) can be represented in the form $\dot{\varphi}=\pm\surd\overline{f(\varphi)}\,  $ and has two branches of solutions corresponding to the different signs.
It can be easily seen that eq. (\ref{14}) is the certain  branch of eq. (\ref{7}) with the proper integration constant.

Thus, a solution of eq. (\ref{14}) $\tilde{\varphi}^{+}(t)$ is a solution of eq. (\ref{7}) but not one of eq. (\ref{9})!

In this sense, quantum theory restores the symmetry broken in classical theory.

The crucial point here is the integration over the functional space $X^{+}$ containing singular functions. If it were the space $C$ we would get the ordinary result obtained after integration over the Wiener measure $\exp\left\{ -\frac{1}{2}\int \limits _{0}^{1}(\dot{\varphi}(t))^{2}dt \right\}\,d\varphi\,.$

Now let us find the explicit form of the solution $\tilde{\varphi}^{+}(t)\,.$ Depending on the bound value  $\tilde{\varphi}^{+}(-T)\equiv-\alpha$ it is
\begin{equation}
   \label{15}
\tilde{\varphi}_{\alpha}^{+}(t)=\beta \tanh(bt+c)\,,\ \ \  for\ -\alpha>-\beta\,,
\end{equation}
or
\begin{equation}
   \label{16}
\tilde{\varphi}_{\alpha}^{+}(t)=\beta \coth(bt+c)\,,\ \ \  for\ -\alpha<-\beta\,.
\end{equation}

If we set the integration constant $c$ to be equal to zero, the solution is the odd function  $\tilde{\varphi}^{+}(-t)=\tilde{\varphi}^{+}(t)\,.$
In this case,
$$
A_{+}\left(\tilde{\varphi}^{+}\right)=A\left(\tilde{\varphi}^{+}\right)\,.
$$

So far, we have considered the theory given by the action $A_{+}\,.$ However, the similar picture holds for the "mirror" action
\begin{equation}
   \label{17}
 A_{-}= A +a \int \limits _{-T}^{+T}\varphi(t)dt-\frac{a}{3}\left[\varphi^{3}(+T)-\varphi^{3}(-T) \right]+a\beta^{2}\left[\varphi(+T)-\varphi(-T) \right]\,.
\end{equation}
The corresponding substitution is
\begin{equation}
   \label{18}
\chi(t)=\varphi(t)-a\int \limits _{-T}^{t}\left(\varphi^{2}(\tau)-\beta^{2}\right)d\tau\,.
\end{equation}
And we have the equality of the functional integrals
\begin{equation}
   \label{19}
\int\limits _{X^{-}}\,F(\varphi)\,\exp\{-A_{-}(\varphi)\}d\varphi=\int\limits _{C[-T,+T]}\,F(\varphi(\chi))\,\exp\left\{ -\frac{1}{2}\int \limits _{-T}^{+T}(\dot{\chi}(t))^{2}dt \right\}\ d\chi \,.
\end{equation}
The structure of the space $X^{-}$ is the same as of the space $X^{+}\,.$ But the functions from $X^{+}$ and the functions from $X^{-}$ have singularities  at different points.

The solutions $\tilde{\varphi}^{-}(t)$ of the equation of motion
\begin{equation}
   \label{20}
\dot{\varphi}(t)+a\left(\varphi^{2}(t)-\beta^{2}\right)=0
\end{equation}
obtained in the classical limit belong to the other branch of the solutions of  eq. (\ref{7}). They are connected with the solutions (\ref{15}), (\ref{16}) by the substitution $t\rightarrow-t\,.$

Now, having in mind the above results one can ask the question: "Do we really understand the Standard Model?".

\end{document}